\documentclass{elsart}
\usepackage{epsf,latexsym,longtable}
\newcommand{\kt}{k_B T}
\newcommand{\rol}{n^*_l}
\newcommand{\rov}{n^*_v}
\newcommand{\usf}{U_{sf}}
\newcommand{\sssr}{\sigma^{*}_{ss}}
\newcommand{\ssfr}{\sigma^{*}_{sf}}
\newcommand{\sss}{\sigma_{ss}}
\newcommand{\ssf}{\sigma_{sf}}
\newcommand{\sff}{\sigma_{ff}}
\newcommand{\esf}{\epsilon_{sf}}
\newcommand{\eff}{\epsilon_{ff}}
\newcommand{\er}{\epsilon_r}
\newcommand{\gsf}{\gamma_{sf}}
\newcommand{\gsl}{\gamma_{sl}}
\newcommand{\gsv}{\gamma_{sv}}
\newcommand{\glv}{\gamma_{lv}}
\newcommand{\gsfr}{\gamma^{*}_{sf}}
\newcommand{\gslr}{\gamma^{*}_{sl}}
\newcommand{\gsvr}{\gamma^{*}_{sv}}
\newcommand{\glvr}{\gamma^{*}_{lv}}
\newcommand{\dgsfde}{\partial \gamma_{sf} / \partial \epsilon_{sf}}

\newcommand{\dgsvde}{\partial \gamma_{sv} / \partial \epsilon_{sf}}
\newcommand{\dgsfder}{\partial \gamma^{*}_{sf} / \partial \epsilon_{r}}
\newcommand{\dgslder}{\partial \gamma^{*}_{sl} / \partial \epsilon_{r}}
\newcommand{\dgsvder}{\partial \gamma^{*}_{sv} / \partial \epsilon_{r}}
\newcommand{\cost}{\cos \theta }
\newcommand{\dcost}{\partial \cos \theta }
\newcommand{\dcde}{\partial \cos \theta / \partial \esf}
\newcommand{\dcder}{\partial \cos \theta / \partial \er}
\begin{document}
\journal{Physica A}
\begin{frontmatter}
\title
{Wetting and drying of a rigid substrate under variation 
of the microscopic details}

\author{C. Bruin}

\address{
Laboratorium voor Technische Natuurkunde, Technische Universiteit Delft\\
P.O. Box 5046, 2600 GA Delft, the Netherlands}


\abstract
Wetting and drying of a rigid substrate by a
Lennard-Jones fluid in molecular dynamics simulations is reported.
The size of the substrate particles, being smaller than the fluid particles
in former simulations, is now taken to be equal to, respectively larger 
than, that of the fluid particles.
Recently, for the latter type of system a first order
drying transition has been reported. 
Like before we find a continuous-like transition for all
systems considered.
This also holds for substrates with incompletely-filled top layers, the
so-called molecularly rough surfaces.
All systems studied behave qualitatively alike, but inconsistencies are found
in the solid-vapour surface tension on approach of the wetting
transition and for the solid-fluid surface tension in general.
\endabstract
\keyword
wetting, drying, molecular dynamics
\endkeyword
\end{frontmatter}

\section{Introduction}

In a recent molecular dynamics (MD) paper of fluid wetting on molecularly
rough surfaces, Tang and Harris \cite{tan1} present also results for the
drying transition on a molecularly smooth surface.
These results suggest a first order drying transition and even a metastable
partially-dry state as a value of $\cost$ significantly smaller than -1 is
reported.
The rigid substrate consists of static particles on an FCC lattice.
The smooth-surface substrate contains three (100) layers.
For the rough surface a partially-filled (100) layer is added on
both sides of the substrate.
For the evaluation of the contact angle $\theta$, Young's relation
has been applied
\cite{row}
\begin {equation}
\cos \theta = (\gsv - \gsl) / \glv,
\label{cosdef}
\end {equation}
where $\gamma_{ij}$ is the surface tension of the interface between
phases $i$ and $j$ and the subscripts $s$, $l$ and $v$ denote the
solid, liquid and vapour phase, respectively.

In a Comment
\cite{comm} we have objected against the
Tang and Harris expression for $\gsf$, where $f$ stands for fluid 
(either liquid or vapour).
Harris has complied with our objection in his Reply
\cite{tan2}.
After correction, the above-mentioned result for
$\cost$ remains and is at variance with results obtained by us
before \cite{bru}.
From our results, we concluded \cite{bru} the drying transition to be 
continuous or very weakly first order.
Because of this discrepancy, we performed MD simulations on the 
Tang and Harris smooth- and rough-surface systems. 
A system with
fluid and solid particles of equal size (in between our original system
and the Tang and Harris system)
has been studied too.

Another reason for reexamining the systems studied by Tang and Harris
\cite{tan1} is the surprising behaviour of $\gsv$, that turns out not to
decrease monotonously with increasing substrate-fluid interaction parameter
$\esf$, even so after correction
\cite{tan2}.
The derivative $\dgsvde$ can be determined directly
in the simulations and was always
found to be negative in our previous simulations.
This derivative has not been reported by Tang and Harris.

We start out in section 2 with a survey of the systems studied.
In section 3 we present results for the contact angle, $\cost$,
and its derivative $\dcder = \eff \dcde$.
The substrate-fluid surface tensions are shown in section 4 and
discrepancy with their derivatives is exposed.
As a check on internal consistence we present
data on the liquid-vapour surface tension, the pressure and
the liquid and the vapour density in section 5.
We end up with a summary of our conclusions in section 6.
In the Appendix Table 3 contains the essential drying/wetting data.

\section {Description of the systems simulated}

The systems we simulate contain both substrate
and fluid particles  that interact through a truncated
and shifted Lennard-Jones (LJ) potential
\begin {equation}
\Phi_{AB} (r) = \left\{ \begin{array}{cc}
                  \phi_{AB} (r) - \phi_{AB} (r_c), & r \leq r_c \\
                           0, & r > r_c
                           \end{array}
                   \right.
\end {equation}
with
$\phi_{AB} (r) = 4 \epsilon_{AB} \{ ( \sigma_{AB} / r )^{12} -
 ( \sigma_{AB} / r )^6 \}$,
where $A$ and $B$ stand for either "substrate" or "fluid", 
$r$ denotes the distance 
between the particles, $\epsilon_{AB}$ sets the energy scale and
$\sigma_{AB}$ the length scale of the potential.
The cut-off radius is $r_c = 2.5 \sigma_{AB}$.
We have used the same program as before
\cite{bru}, although the "massive stochastic collisions" method of
Tang and Harris \cite{tan1} has been checked as well.
For our expressions of the surface tensions $\gsf$ and $\glv$, as 
integrals over pressure tensor components, we refer to Nijmeijer et al.
\cite{nijm3,nijm4}. For the relation between the derivative $\dgsfde$ and
the potential energy $\usf$ of the substrate-fluid interaction we refer to
references
\cite{bru,nijm4,swol86,hend92}, $\usf$ is calculated for each side
of the substrate separately.

\begin{table}
\caption{
Summary of the systems simulated for $\ssfr = 1.0$ and
$\ssfr = 1.1$ for smooth (sm) substrates, and for rough substrates for
$\ssfr = 1.1$ with coverage of the top layers 1/4 (r1), 2/4 (r2) and
3/4 (r3) respectively.
$N_c$ denotes the number of unit cells of the substrate
both in the $x$- and $y$-direction and the third column gives the resulting
linear dimensions ( $L_x^* = L_y^*$ ).
The next entries give the length of the system $L_z^*$ perpendicular
to the substrate,
the number of substrate particles ($N_s$)
and the number of fluid particles ($N_f$) respectively.
The number of time steps per run is given: $\Delta t$/run, where the number of
runs ranges from fifty up to eight hundred (for slowly equilibrating systems).
The liquid-vapour surface tension is given in the last column.
}
\begin{tabular}{lrrrrrrc}
\hline
system &$N_c$ &$L_x^*=L_y^*$ &$L_z^*$ &$N_s$ &$N_f$ &$\Delta t/run$ &$\glvr$\\
\hline
sm     &      &              &        &      &      &       &$\ssfr = 1.0$\\
\hline
S3    & 5  & 7.94  &38     &  150  &  816  & 60000 &0.239(2) \\
S2    &10  &15.87  &38     &  600  & 3264  & 15000 &0.231(1) \\
S1    &20  &31.75  &38     & 2400  &13056  &  4000 &0.227(3) \\
\hline
sm     &      &              &        &      &      &       &$\ssfr = 1.1$\\
\hline
S3    & 4  & 7.62  &38     &   96  &  752  & 60000 &0.243(2) \\
S2    & 8  &15.24  &38     &  384  & 3008  & 15000 &0.231(1) \\
S1    &16  &30.48  &38     & 1536  &12032  &  4000 &0.229(3) \\
\hline
r1     &      &              &        &      &      &       &$\ssfr = 1.1$\\
\hline
S2    & 8  &15.24  &41     &  448  & 3008  & 15000 &0.233(1) \\
S1    &16  &30.48  &41     & 1792  &12032  &  4000 &0.232(2) \\
\hline
r2     &      &              &        &      &      &       &$\ssfr = 1.1$\\
\hline
S2    & 8  &15.24  &41     &  512  & 3008  & 15000 &0.231(1) \\
S1    &16  &30.48  &41     & 2048  &12032  &  4000 &0.228(4) \\
\hline
r3     &      &              &        &      &      &       &$\ssfr = 1.1$\\
\hline
S2    & 8  &15.24  &41     &  576  & 3008  & 15000 &0.229(1) \\
S1    &16  &30.48  &41     & 2304  &12032  &  4000 &0.222(6) \\
\hline
\end{tabular}
\end{table}
As before
\cite{bru}, all systems have been studied at a constant temperature 
$T^* = \kt / \eff = 0.9$, that was kept fixed by scaling the velocities of the
fluid particles every hundredth time step.
For the integration of the equations of motion the leap-frog algorithm 
has again been used with a reduced time step
$\Delta t^* = \Delta t \sqrt{\eff} / ( \sff \sqrt{m_f} ) = 0.01$, where $m_f$
denotes the mass of a fluid particle.
Every twentieth time step a configuration is used for the calculation of the
surface tensions and the substrate-fluid potential energies, 
as well as the density profiles.
The key parameter determining drying/wetting behaviour is the
interaction parameter $\esf$ in units of $\eff: \er = \esf / \eff$.

In Table 1 the characteristics of the various systems are given.
System S3 for $\ssfr = \ssf / \sff = 1.1$, labeled "$\sigma$11-sm", is
identical to the smooth-surface system of 
Tang and Harris
\cite{tan1,tan2}, except for the length $L^*_z$ of the system.
From their specifications $\sssr = 1.2$ and $L^*_x =L^*_y = 7.62$,
with 4 unit cells lateral in the substrate, we derive the
nearest-neighbour distance in the substrate to be the position of the
minimum of the LJ potential.
It turns out that these interface dimensions are rather close to those of
the system S3 for $\ssfr = 0.941$ ($\sssr = 0.833$), with 6 lateral unit 
cells, in our former paper \cite{bru}: $L^*_x =L^*_y = 7.94$.
The dimensions of the other smooth-surface system mentioned in Table 1,
$\ssfr = 1$ ($\sssr = 1$) with 5 lateral unit cells, are identical:
$L^*_x =L^*_y = 7.94$.
Therefore we denote all three systems with "S3", to specify the size
of the interface area.
As before, it is necessary to study larger interface area's as well,
where S2 has a 4 times larger and S1 has a 16 times larger surface area
(see also \cite{bru}).
For the rough-surface systems of Tang and Harris, with incompletely-filled
extra top layers on both sides of the substrate, only our results for 
the larger systems are mentioned.
For the different system sizes the number of time steps per run is 
such that the statistical accuracy and the CPU time per run
are about constant.
For the last column of Table 1 the liquid-vapour surface tension 
$\glv^{*} = \glv \sff^2 / \eff$
has been averaged over all $\er$ values for the pertaining system.
The numbers between parentheses denote the uncertainty
(one standard deviation) in the last one or two digits.

\section {Results of the simulations for {\boldmath $\cost$}}

The results of the simulations, regarding $\gslr$, $\gsvr$ and $\cost$ and
their derivatives with respect to $\er$, or rather $\esf$, are listed in the
Table 3 in the Appendix.
In Fig. 1, $\cost$ for three smooth-surface systems and one 
rough-surface system is shown.
The results of the two omitted rough-surface systems $\sigma11$-r1 and 
$\sigma11$-r3 are rather 
close to those of the smooth-surface system $\sigma11$-sm.
Our earlier results for $\ssfr = 0.941$ ($\sigma94$) are represented
by a thick full line. It is taken from 
\cite{bru}, where its origin is explained (multi-histogram method
for $\er \leq 0.25$ and
cubic spline fit for $\er > 0.25$) and it is shown
to represent very well the various systems reported there.
For completeness two data around the wetting transition have been added 
(circles, system S33 \cite{bru}, that has a liquid phase 
doubled in the $z$-direction with respect to S3).
Comparing the various systems
we conclude that variation of the microscopic details of the substrate
does not effect the qualitative behaviour of drying 
and wetting, as expected.
For the smooth-surface systems, increasing $\ssf$ (and thereby $\sss$) 
gives the same behaviour at larger values of $\er$.
The external potential, that the substrate imposes on the fluid, can be
approximated by summing over all substrate particles and averaging over the
directions parallel to the surface. This external potential
gets less deep for increasing $\sss$, due to the lower density of the
substrate, but it has a larger range due to a larger $\ssf$.
From the integrated
negative part of the external potentials,
scaling factors can be deduced for
$\er$, 1.24 for $\sigma1$ and 1.36 for $\sigma11$-sm,
with respect to $\sigma94$.
Both scaled results, of $\sigma94$, are shown in Fig. 1 
by thick full lines too.

\begin{figure}
\epsfxsize=13.5cm
\epsfbox{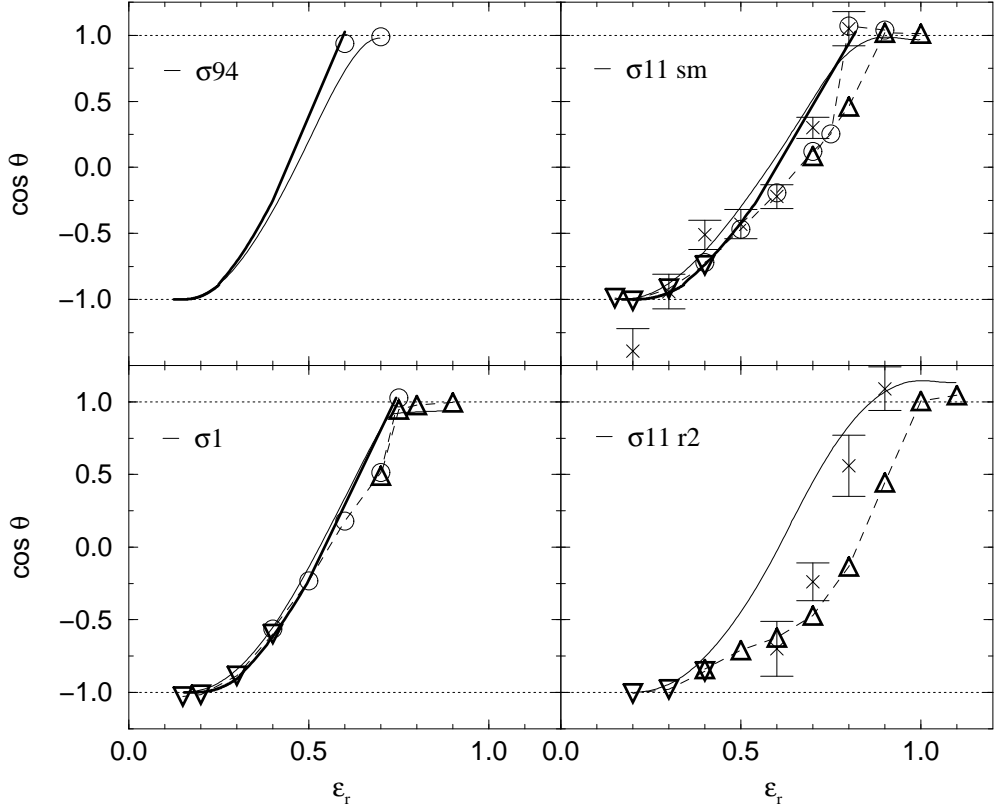}
\caption{$\cost$ as a function of $\er$.
Dotted lines at 1 and -1 denote the completely-wet and completely-dry
values of $\cost$ respectively.
Results for the various system sizes are denoted by O (S3),
$\bigtriangleup$ (S2) and $\bigtriangledown$ (S1). Error bars have
been omitted as they are smaller than the symbol size.
Dashed lines connecting the symbols should guide the eye.
Thin full lines result from integrating spline fits, to $\dcder$ (Fig.2).
Thick full lines represent the results for $\sigma94$ \protect\cite{bru},
and the
scaled versions (see text) for $\sigma1$ and $\sigma11$-sm respectively.
The corrected Tang and Harris data (crosses) have error bars
of two standard deviations.}
\end{figure}
For reasons we will discuss later, we have exploited another route as well
to obtain $\cost$ ( and $\gslr$ and $\gsvr$).
As we have calculated also $\dgslder$ and $\dgsvder$, we 
have made spline fits for these quantities and for $\dcder$ (see also Fig.2)
and have integrated these fits.
For the systems mentioned in Table 1 we have taken together data for the
various surface area's, in case of overlapping data we always dropped 
the data belonging to smaller surface area's.  
For system $\sigma94$ we took the data of D1,
special purpose computer system of size S1 \cite{bru}, extended with
two data points of S33 ($\er = 0.6$ and 0.7).
The results from the integration of the spline fits to $\dcder$ are shown
by the thin full lines in Fig.1, where a trivial integration constant
of -1 has been added.
Integrating the fits to $\dgslder$ and $\dgsvder$ (see section 4) 
to obtain $\gslr$ and $\gsvr$
and combining the resulting surface tensions according to eq.(1) 
gives the same curves.
The integration constants are less trivial here and we preferred to
adjust the surface tensions at the lowest value of $\er$ (see Fig.3 and 4).
For $\sigma11$-sm in Fig.1 the thin full line is in much better agreement
with the thick full line than the data points are.
In general, we can conclude that this procedure of scaling $\er$ accounts
for the main effect of the increase of $\ssf$, and $\sss$.

A striking difference between Harris's data and ours (for $\sigma11$-sm)
is at $\er = 0.2$, where he obtains $\cost = -1.39(17)$, well below the
completely-dry value of -1, the error bar being two standard deviations.
This value is obtained with a $\glv^* = 0.191$, which is much lower than
the value 0.234(2) we obtained for this system size (S3,\cite{bru}).
Substituting the latter value of $\glv^*$, one obtains $\cost = -1.14$ 
that is still too low, but that does not rule out the value -1 in 
view of the error bar.
Another difference is that Harris finds an incompletely-dry state 
for $\er = 0.3$ for the system size S3. We also simulated S3 for $\er = 0.3$
and found the liquid phase to detach from the substrate, like we found for 
S3 ($\sigma94$,\cite{bru}) for $\er = 0.275$.
For this reason we studied a 16 times larger surface area (S1) for
$\er \leq 0.4$.
The reason for the above-mentioned discrepancy lies in the different way
of keeping the temperature constant. We also tried Tang and Harris's method 
of "massive stochastic collisions" for system S3 ($\sigma11$-sm) and found
$\cost = -1.00(2)$ for $\er = 0.2$.
However, both the liquid density $\rol$ and $\glv^*$ become too small and 
the vapour density $\rov$ becomes too large, as Tang and Harris find too
\cite{tan1}.
In our opinion this originates from the fact that the tendency of the liquid
phase to detach from the substrate is frustrated by assigning "random" 
velocities to the fluid particles every 500th time step.

\begin{figure}
\epsfxsize=13.5cm
\epsfbox{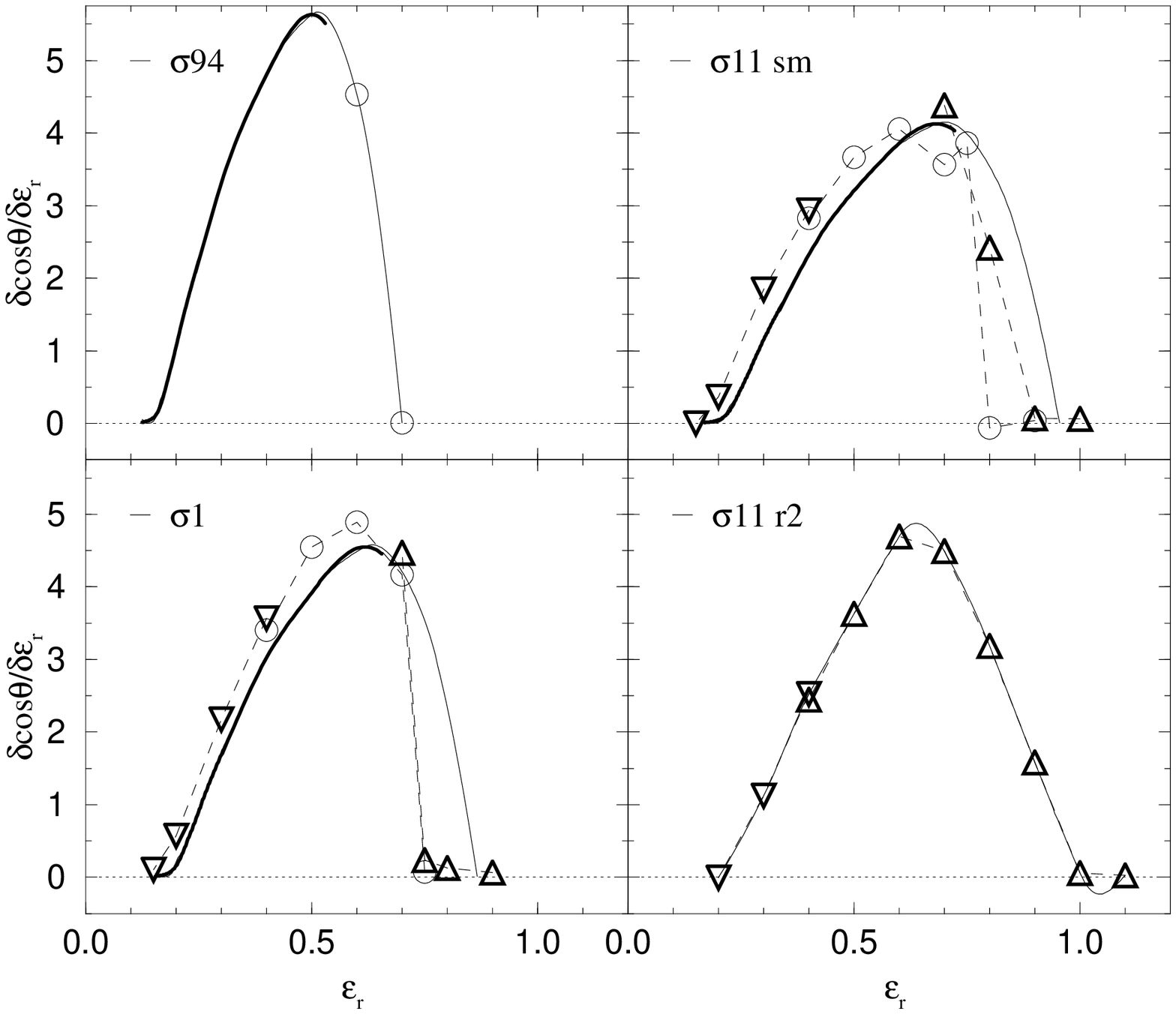}
\caption{$\dcder$ as a function of $\er$ for the same systems depicted
in Fig.1. The dotted line at zero
gives both the completely-dry and the completely-wet value of $\dcder$.
For $\sigma94$ the thick full line represents the results and the
thin full line is the spline fit, see Fig.1.
Both lines have been scaled for $\sigma1$ and $\sigma11$-sm.
For $\sigma11$-r2 the full line represents its spline fit.
}
\end{figure}
Another interesting feature shown in Fig.1 is the size dependence of the 
wetting transition for $\sigma11$-sm. System S3 becomes completely-wet
for $\er = 0.8$ (in accordance with Harris's result), where for S2
$\er = 0.9$ applies.
For $\sigma1$ this size dependence is marginal or absent ($\er = 0.75$) 
and for $\sigma94$ we found no indication for it at all \cite{bru}.
For the rough-surface systems ($\sigma11$), the size dependence is 
present as well,
and even more pronounced; because of the behaviour of $\gsvr$ (to be discussed
later), we did not pursue the simulations of S3 for the
rough substrates.

In Fig. 2 the derivative $\dcder$ is shown for the same simulations 
depicted in Fig. 1.
The size dependence of the wetting transition for $\sigma11$-sm manifests
itself very clearly.
For system size S3, $\dcder$ falls from 4 to 0 (the completely-wet value)
going from $\er = 0.75$ to $\er = 0.8$.
System size S2 gets completely wet for $\er = 0.9$, see also Fig.1.
The value of $\dcder$ at $\er = 0.8$ looks suspicious, lying in between 
the top value and the completely-wet value zero.
Indeed of the probes for equilibrium, $\dgsvder$ and $\dgslder$, 
the first one shows considerable fluctuations with time, 
as we also found for other systems
on approach of the wetting transition.
Therefore we followed this system for 5.25x$10^6$ time steps after an
equilibration time of 7.5x$10^5$ time steps.
For comparison: S2 at $\er = 0.9$ became completely wet after 4.5x$10^6$
time steps and was followed for 1.5x$10^6$ time steps afterwards.
The scaling of $\er$ for $\sigma94$ to $\sigma1$ and to $\sigma11$-sm, 
mentioned above for $\cost$, should be accompanied here with scaling 
down $\dcder$ with the same factor.
The spline fit to $\dcder$ for $\sigma94$ (using the same systems as for
$\cost$) is shown by the thin full line,
only visible for $\er > 0.5$.
This fit has been scaled as well into results for $\sigma1$ and
$\sigma11$-sm, these are also represented by thin full lines.
From the comparison of the scaled data with the measured data 
for $\sigma1$
and $\sigma11$-sm we conclude that this scaling gives a reasonable 
estimate of the effect of increasing $\ssf$, and $\sss$.
It should be noted that the thin full line for $\sigma11$-r2 in Fig.2
represents the spline fit to the data and this fit has been integrated to
obtain $\cost$ in Fig.1 for $\sigma11$-r2.
\begin{figure}
\epsfxsize=13.5cm
\epsfbox{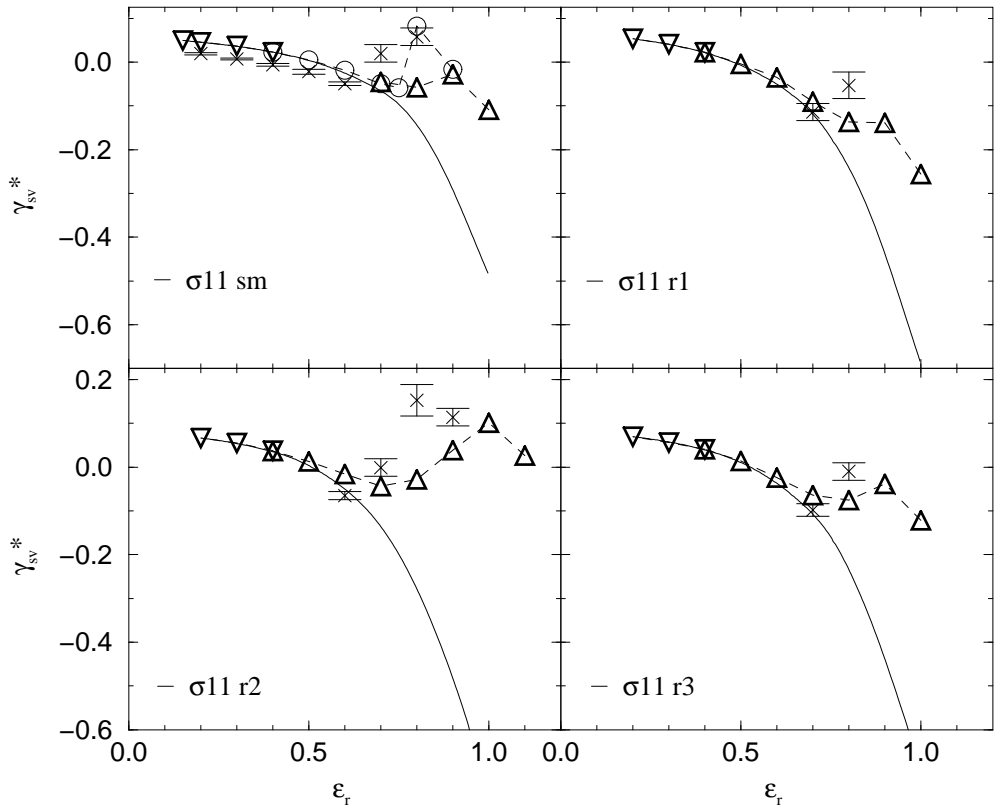}
\caption{$\gsvr$ for the various systems with $\ssfr = 1.1$.
Results for the various system sizes are denoted by O (S3),
$\bigtriangleup$ (S2) and $\bigtriangledown$ (S1). Error bars have
been omitted as they are smaller than the symbol size.
Dashed lines connecting the symbols should guide the eye.
The corrected Tang and Harris data (crosses) have error bars
of two standard deviations.
Full lines result from integrating the spline fits 
to $\dgsvder$ (see Fig.6).}
\end{figure}

\section {Surface tensions}
In Fig.3 the results of $\gsvr$ for the various systems with
$\ssfr = 1.1$ are shown.
The data of Harris \cite{tan2} all show an increase of $\gsvr$
with increasing $\er$ on approach of the wetting transition 
(system size S3).
In our previous simulations \cite{bru} for $\sigma94$ we never 
encountered such a behaviour for any system size.

\begin{figure}
\epsfxsize=13.5cm
\epsfbox{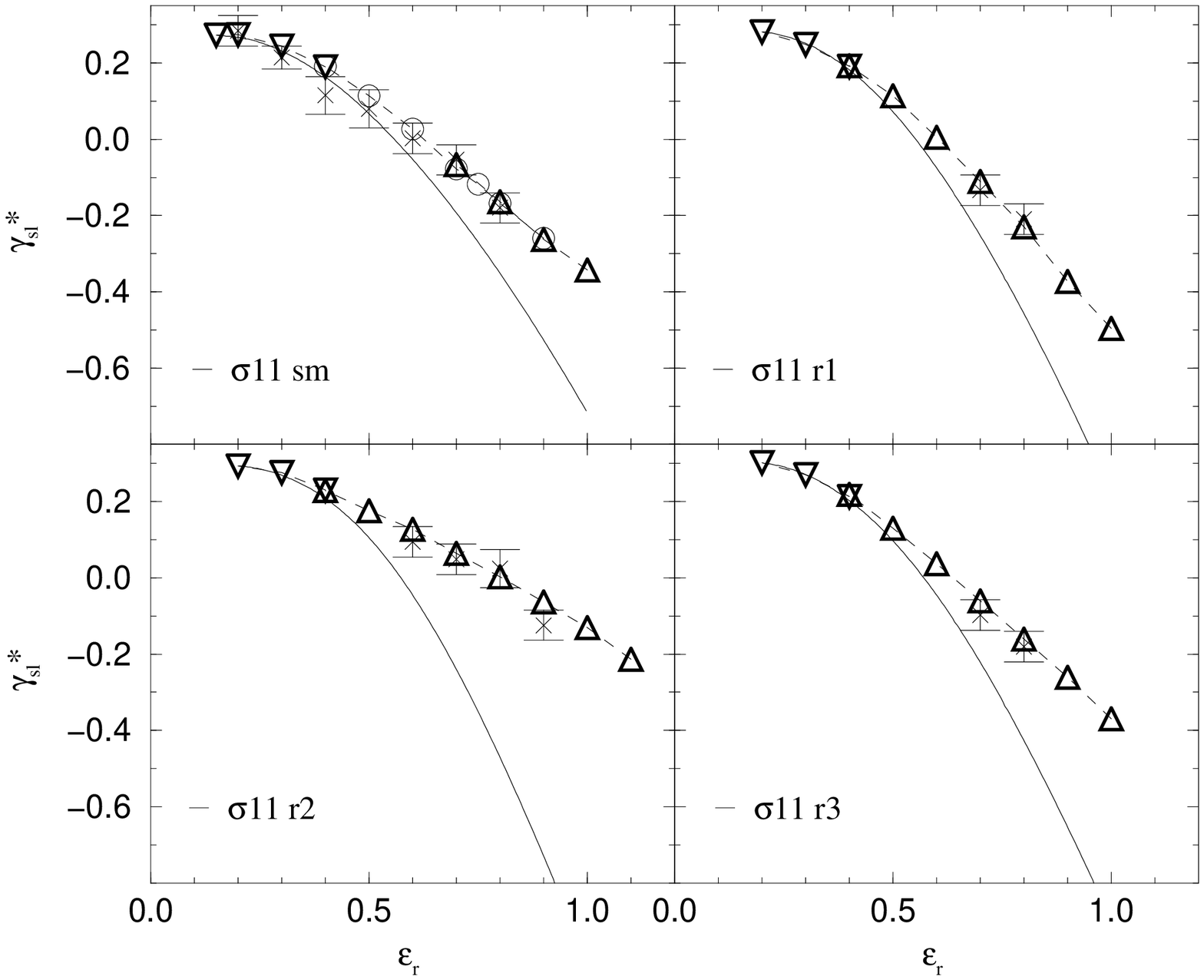}
\caption{$\gslr$ for the various systems with $\ssfr = 1.1$
(see Fig.3).
Full lines result from integrating the spline fits to $\dgslder$ 
(see Fig.6).
}
\end{figure}
For the smooth-surface system ($\sigma11$-sm), behaviour comparable 
to that of the data of Harris is shown by
system S3, although $\gsvr$ is decreasing up to $\er =0.75$.
For $\er =0.8$, the same completely-wet state is found.
The increase of $\gsvr$ for S2, with a larger interface area,
is less prominent ($\er =0.9$), but does exist.
It looks like the system jumps to the completely-wet branch, that is really
$\gslr+\glvr$, at a too low value of $\er$.
So we performed simulations for S1 as well in this region, but we still found
the tendency of $\gsvr$ to increase, although the exceptional 
long equilibration times prevented us from arriving at a fully equilibrated
completely-wet state for such a large system.
From the rough-surface systems, $\sigma11$-r1 and $\sigma11$-r3 show 
very much the same
behaviour as the smooth-surface system, which also holds for the few
data of Harris.
The situation for $\sigma11$-r2, that becomes completely wet at $\er = 1$,
looks different as the increase of $\gsvr$ takes place over
a much larger range: $0.7 < \er < 1$.
The increase of $\gsvr$ just before the wetting transition,
and for $\sigma11$-r2 in a larger region, is in contradiction 
with $\dgsvder$ as depicted in Fig.6, being always negative and getting more
negative for increasing $\er$.
We therefore made spline fits to this derivative (see Fig.6).
Integration of the spline fits, taking the integration constant such
that values at low $\er$ match the directly measured data, results
in the full lines of Fig.3.
The agreement up to $\er \approx 0.6$ is good, but it gets dramatically bad
for higher values of $\er$.

\begin{figure}
\epsfxsize=13.5cm
\epsfbox{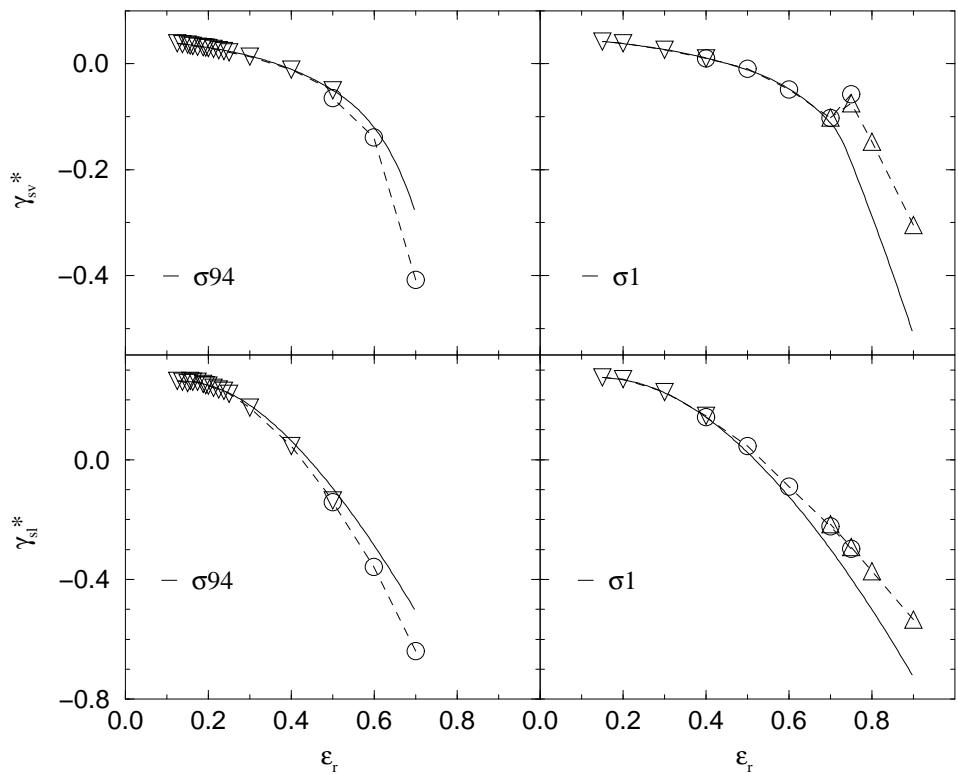}
\caption{
$\gsvr$ for the smooth-surface systems $\sigma94$ 
and $\sigma1$ (two upper graphs), and $\gslr$ (two lower graphs).
Full lines denote the integrated spline fits of the respective 
derivatives.
Symbols for $\sigma1$ are according to Fig.3.
For $\sigma94$ data of D1 ($\bigtriangledown$) and S33 (circles) are given.
}
\end{figure}
In Fig.4 $\gslr$ is shown for the same systems.
There is very good agreement with the data of Harris (crosses).
The data look much more regular here, and of course nothing special
is happening to $\gslr$ on approach of the wetting transition.
However, applying the same procedure of integrating now the spline
fits to $\dgslder$ (Fig.6), resulting
in the full lines, also shows a large discrepancy with the directly
measured surface tensions.
The discrepancy is even starting at lower values of $\er$.
One should note that at these low values of $\er$, $\gslr$
is affected by the build-up of the liquid phase near the substrate
(for increasing $\er$).
The same phenomenon applies to $\gsvr$ on approach of the
wetting transition.
We calculated the two contributions to $\gsfr$ (referring to
both $\gsvr$ and $\gslr$) separately, i.e. 
the fluid-fluid and the substrate-fluid part.
They have opposite signs and their absolute values become five to ten times
larger for $\gsvr$ than the quantity itself in the completely-wet
region.
However, no suspect behaviour could be found for the constituents.
A possible reason, for the discrepancy between 
$\gsfr$ and its derivative $\dgsfder$,
is registration of the fluid particles onto the
substrate, that is excluded for system $\sigma94$.
We tried to check this conjecture by visual inspection of the 
first few liquid layers nearest to the substrate, but did not find 
evidence for it. 
A quantitative analysis, through a pair distribution function or
a (two-dimensional) structure factor, should be decisive.

\begin{figure}
\epsfxsize=13.5cm
\epsfbox{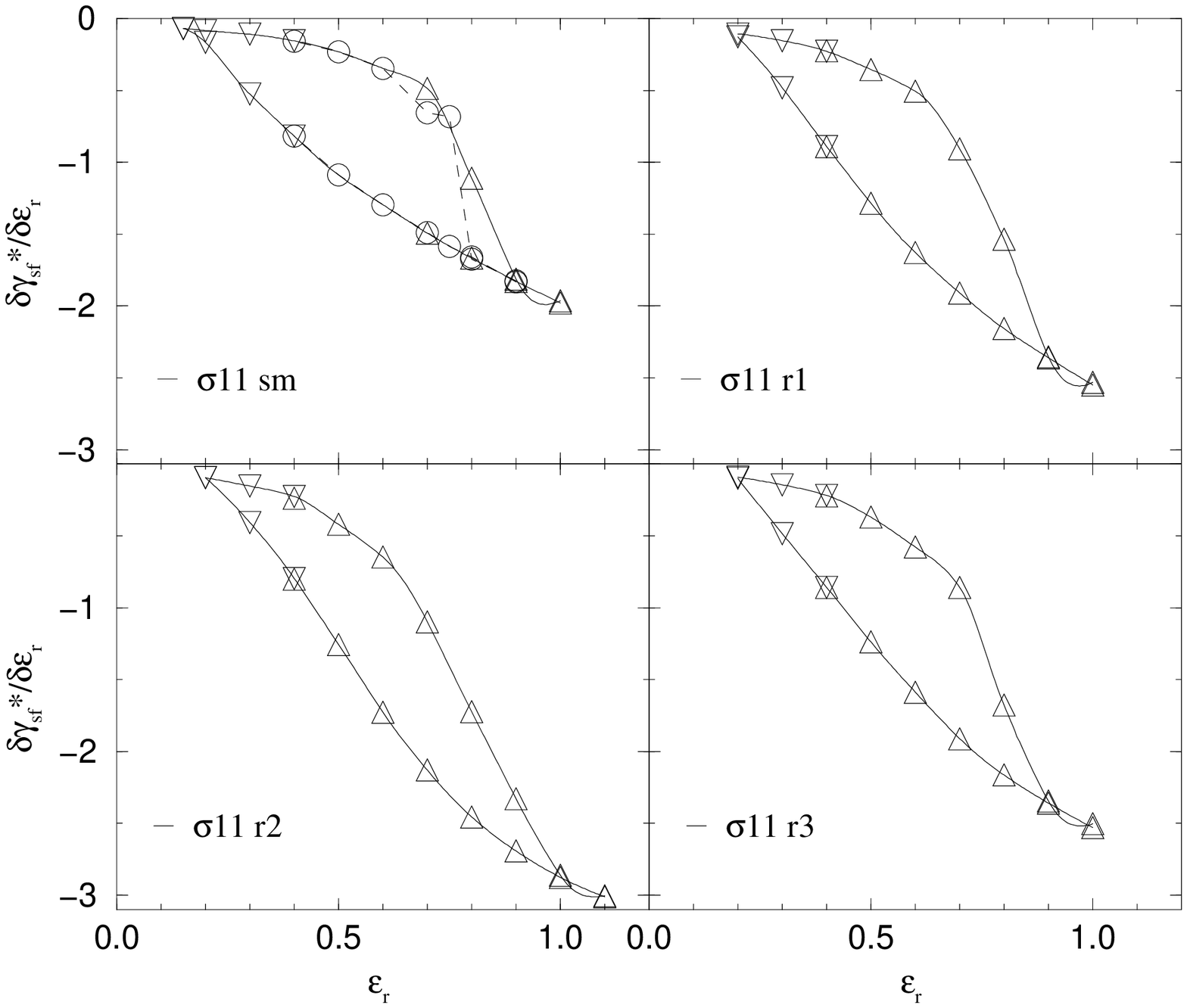}
\caption{$\dgsvder$, upper curves, and $\dgslder$, lower curves,
for the various systems with $\ssfr = 1.1$ (see Fig.3 and 4).
Full lines show the spline fits to the data, that have been integrated
to obtain $\gsvr$ and $\gslr$ (also full lines)
in Fig.3 and Fig.4 respectively.}
\end{figure}
We observe the discrepancy growing, with the substrate becoming 
less smooth, as it does when the substrate particles become larger.
For $\sigma94$ \cite{bru} we find a small discrepancy (see Fig.5), 
that has opposite
sign with respect to the discrepancies displayed in Fig.3 and 4.
For $\sigma1$ the particle sizes are equal and we find a relatively small
discrepancy (see Fig.5).
For $\sigma11$-sm the substrate particles are larger than 
the fluid particles, and the discrepancy is large (Fig.3 and 4).
For $\sigma11$-r1 and for $\sigma11$-r3 the discrepancy is about the same
as for $\sigma11$-sm.
Apparently both systems, with a top layer of the substrate consisting 
for one quarter out of particles and for one quarter out of vacancies 
respectively, do not look rougher to the fluid than $\sigma11$-sm does.
One could argue that $\sigma11$-r2, with a 50\% filled top layer,
looks rougher than $\sigma11$-sm and can be supposed to favour
registration even more, the discrepancy is 
larger indeed (see Fig.3 and 4).

In Fig.6 the derivatives $\dgsvder$ and $\dgslder$ are shown, together
with the spline fits.
The different systems show very much the same behaviour.
It should be noted however that, considering systems $\sigma11$-sm,
$\sigma11$-r3, $\sigma11$-r1 and $\sigma11$-r2, and
in that order, $\dgsvder$ (upper curves) starts 
to decrease more
rapidly at lower values of $\er$ and shows a smoother overall behaviour.

\section {Liquid-vapour equilibrium}

As there are always coexisting liquid and vapour phases in our simulations
with at least one interface in between, we obtained quite some data
on the liquid density $\rol$, the vapour density $\rov$ and the pressure
$p^*$ at coexistence ($T^* = 0.9$), as well as on the liquid-vapour
surface tension $\glvr$.
As a check on the consistence of our simulations and possibly as a reference,
the numbers are given in Table 2.

In Table 1 we already presented values for $\glvr$ for the various systems.
We average these values for the various system sizes separately
and find $\glvr$ to depend on the size of the surface area (Table 2), 
like we reported before \cite{bru}, where the value
for S3 (0.236) differs slightly from the one mentioned in \cite{bru}
(0.234) because of other weight factors.
For completeness we mention the result for a system (S0) with a 4 times larger
interface area \cite{bru} than S1: $\glvr = 0.227(1)$, that fits in
quite nicely with the results obtained here.
Two other relatively recent results are from Adams and Henderson \cite{adams}
for a system with a bit smaller interface area than S3: $\glvr = 0.23(1)$,
and Haye and Bruin \cite{haye}:
$\glvr = 0.223(3)$ {S1}.
For the calculation of $\cost$ with eq.(1) we used the same values of
$\glvr$ as before \cite{bru}.
The discrepancy of a few percent between the old value for S3 and 
the present one has been ignored.

For the pressure our present results suggest size dependence as well,
as it is increasing with system size; the error bars
prevent us from drawing this conclusion.
Moreover, it is not supported by the data of our previous work.
For system size S0 we obtained $p^*$ = 0.03130(1),
Adams and Henderson \cite{adams} report: $p^*$ = 0.0322(13) and Haye 
and Bruin \cite{haye}: $p^*$ = 0.0308(2).

\begin{table}
\caption{
Liquid-vapour coexistence data at $T^* = 0.9$ for the various
system sizes.
Each row with present results is followed by one with previous
results \protect\cite{bru} for the same quantity.
The last column contains the average value, where appropriate.
In the average value for $\rol$ \protect\cite{bru} only the results of the
doubled liquid phase systems S33, S22 and S11 have been taken into
account (see text).
}
\begin{tabular}{lllllll}
\hline
            & S3       & S2       & S1       & D1       & $< >$ \\
\hline
$\glv^*$    &0.241(2)  &0.231(1)  &0.228(2)  &          &          \\
\cite{bru}  &0.236(2)  &0.229(3)  &0.226(1)  &0.225(1)  &          \\
\hline
$p^*$       &0.03109(8)&0.03122(4)&0.03138(7)&          &0.03124(3)\\
\cite{bru}  &0.0313(2) &0.0313(1) &0.03136(6)&0.03102(5)&0.03122(3)\\
\hline
$\rol$      &0.6646(2) &0.6643(1) &0.6645(1) &          &0.6644(1)\\
\cite{bru}  &0.6635(4) &0.6631(5) &0.6631(5) &0.6622(1) &0.6645(1)\\
\hline
$\rov$      &0.0446(2) &0.0449(1) &0.0450(2) &          &0.0449(1)\\
\cite{bru}  &0.0452(3) &0.0453(3) &0.0453(2) &0.0457(1) &0.0452(1)\\
\hline
\end{tabular}
\end{table}
Size dependence can be excluded for the coexisting densities.
We note quite a discrepancy with our previous work for the liquid densities.
System D1 shows the largest discrepancy. In the old results for
S3, S2, and S1, systems with a doubled liquid phase
(S33, S22 and S11 see ref. \cite{bru} ) are included and the numbers are
in between that for D1 and the present results.
The average for $\rol$ over S33, S22 and S11 alone 
is given in the last column and
is in good agreement with the present result.
Since we took narrower boundaries for the determination of the liquid density
in the present simulations
and since more
fluid particles resided in the liquid phase than for D1, S3, S2, S1 and S0
in our previous simulations, we can understand the different outcomes.
Furthermore Adams and Henderson (see above) report:
$\rol$ = 0.671(1) and $\rov$ = 0.040(3) and Haye and Bruin (see above):
$\rol$ = 0.6619(4) and $\rov$ = 0.0454(4).
For the liquid density these results are much higher respectively much lower
than our present result.
We note that for $<\rov> = 0.0452$ in the last row, the S0 result
($\rov = 0.0450(1)$) has been taken into account as well.
The measured value for the temperature, kept fixed only approximately,
equals the prescribed value of 0.9 with an accuracy that is better by
a few decimals than the accuracies of the quantities mentioned above.

\section {Conclusions}

Changing the microscopic details of the substrate does not change
the qualitative behaviour of drying and wetting.
The quantitative change of drying and wetting for the smooth substrates,
due to enlarging the size of the substrate particles,
can be described reasonably well for $\cost$, as it is calculated 
from its derivative, by matching the (approximate) 
external potentials.

The substrate-vapour surface free energy increases at, or just before,
the wetting transition which seems unphysical.
Discrepancy is found between the behaviour of both
$\gsvr$ and $\gslr$ and their derivatives.
The discrepancy gets larger for larger substrate particles and for
more roughness of the substrate.
A conjecture, that registration of the fluid particles
onto the substrate is responsible for this discrepancy,
could not be confirmed by visual inspection and still
awaits a quantitative analysis (pair distribution or structure factor).

Dependence of the liquid-vapour surface tension on the size of the
interface area has been established with better accuracy than before.
Accurate numbers have been found for the coexisting vapour and liquid
densities and the pressure ($T^* = 0.9$), also ensuring internal 
consistence of the simulations.

\begin{ack}
I thank Marco Nijmeijer for his inspiring interest in the present
work and for his fruitful collaboration in the project,
started by Hans van Leeuwen.
\end{ack}

\section*{Appendix}
\setlength{\LTcapwidth}{5.3in}
\begin{longtable}{ccccccc}

  0.90 &-0.534(5) &-0.304(4) &-1.00(3) &-2.370(3) &-2.355(14) & 0.07(13)\kill

\caption{
Surface tension results for the systems with a smooth substrate (sm),
$\ssfr = 1.0$ and $\ssfr = 1.1$, and with a rough substrate (r1, r2 and r3)
for
$\ssfr = 1.1$, with coverage 1/4, 2/4 and 3/4 respectively. In the last 
column $\partial \cost$ is a short-hand notation for
$\dcder$.
}\\
\endfirsthead\\

\caption[]{continued}\\
\hline
$\er   $ &$\gslr $ &$\gsvr $ &$\cost $ &$\dgslder $ &$\dgsvder $ &$\dcost$\\
\hline
\endhead\\

\hline
$\er   $ &$\gslr $ &$\gsvr $ &$\cost $ &$\dgslder $ &$\dgsvder $ &$\dcost$\\
\hline
S2 &$\ssfr=1$ &sm\\
\hline
  0.90 &-0.534(5) &-0.304(4)  & 1.00(3)  &-2.370(3)  &-2.355( 4)  & 0.07( 3)\\
  0.80 &-0.372(4) &-0.148(3)  & 0.98(2)  &-2.130(3)  &-2.102( 3)  & 0.12( 2)\\
  0.75 &-0.293(3) &-0.075(2)  & 0.95(2)  &-2.005(2)  &-1.951( 3)  & 0.23( 2)\\
  0.70 &-0.215(3) &-0.101(1)  & 0.49(1)  &-1.880(2)  &-0.852( 9)  & 4.47( 4)\\
\hline
S3 &$\ssfr=1$ &sm\\
\hline
  0.75 &-0.298(4) &-0.058(4)  & 1.03(2)  &-2.005(4)  &-1.987( 5)  & 0.08( 3)\\
  0.70 &-0.222(4) &-0.102(2)  & 0.51(2)  &-1.870(3)  &-0.896(24)  & 4.16(11)\\
  0.60 &-0.091(6) &-0.048(2)  & 0.18(3)  &-1.612(6)  &-0.467(19)  & 4.89( 9)\\
  0.50 & 0.045(6) &-0.010(1)  &-0.23(3)  &-1.325(5)  &-0.261( 8)  & 4.55( 4)\\
  0.40 & 0.142(7) & 0.010(1)  &-0.56(3)  &-0.989(6)  &-0.192( 7)  & 3.41( 4)\\
\hline
S1 &$\ssfr=1$ &sm\\
\hline
  0.40 & 0.146(5) & 0.010(1)  &-0.60(2)  &-0.994(5)  &-0.188( 2)  & 3.57( 2)\\
  0.30 & 0.226(6) & 0.026(1)  &-0.88(3)  &-0.622(5)  &-0.131( 2)  & 2.17( 3)\\
  0.20 & 0.269(5) & 0.039(1)  &-1.02(2)  &-0.220(6)  &-0.092( 1)  & 0.57( 3)\\
  0.15 & 0.275(6) & 0.042(1)  &-1.03(3)  &-0.098(3)  &-0.075( 1)  & 0.10( 2)\\
\hline
S2 &$\ssfr=1.1$ &sm\\
\hline
  1.00 &-0.342(4) &-0.109(3)  & 1.01(2)  &-1.978(2)  &-1.964( 2)  & 0.06( 2)\\
  0.90 &-0.262(4) &-0.028(3)  & 1.02(2)  &-1.827(2)  &-1.811( 3)  & 0.07( 1)\\
  0.80 &-0.164(2) &-0.058(1)  & 0.46(1)  &-1.667(1)  &-1.110( 9)  & 2.42( 4)\\
  0.70 &-0.066(3) &-0.045(1)  & 0.09(1)  &-1.492(2)  &-0.485( 5)  & 4.38( 3)\\
\hline
S3 &$\ssfr=1.1$ &sm\\
\hline
  0.90 &-0.259(4) &-0.016(4)  & 1.04(3)  &-1.833(3)  &-1.823( 3)  & 0.04( 2)\\
  0.80 &-0.168(4) & 0.082(5)  & 1.07(3)  &-1.659(3)  &-1.673( 3)  &-0.06( 2)\\
  0.75 &-0.118(4) &-0.059(2)  & 0.25(2)  &-1.585(2)  &-0.682(21)  & 3.86( 9)\\
  0.70 &-0.078(6) &-0.049(2)  & 0.12(3)  &-1.490(3)  &-0.656(25)  & 3.57(11)\\
  0.60 & 0.027(5) &-0.018(1)  &-0.19(2)  &-1.295(3)  &-0.345(13)  & 4.06( 6)\\
  0.50 & 0.114(5) & 0.005(1)  &-0.47(2)  &-1.089(4)  &-0.231( 7)  & 3.67( 3)\\
  0.40 & 0.191(5) & 0.023(1)  &-0.72(2)  &-0.817(7)  &-0.157( 3)  & 2.82( 3)\\
\hline
S1 &$\ssfr=1.1$ &sm\\
\hline
  0.40 & 0.190(6) & 0.022(1)  &-0.74(2)  &-0.818(4)  &-0.154( 2)  & 2.93( 2)\\
  0.30 & 0.244(6) & 0.037(1)  &-0.92(3)  &-0.526(6)  &-0.109( 2)  & 1.84( 3)\\
  0.20 & 0.274(6) & 0.046(1)  &-1.01(3)  &-0.165(7)  &-0.082( 1)  & 0.37( 3)\\
  0.15 & 0.273(4) & 0.049(1)  &-0.99(2)  &-0.068(1)  &-0.068( 1)  & 0.00( 1)\\
\hline
S2 &$\ssfr=1.1$ &r1\\
\hline
  1.00 &-0.495(5) &-0.256(5)  & 1.04(3)  &-2.549(3)  &-2.532( 3)  & 0.07( 2)\\
  0.90 &-0.372(5) &-0.139(4)  & 1.02(3)  &-2.360(3)  &-2.352( 3)  & 0.03( 2)\\
  0.80 &-0.229(3) &-0.137(1)  & 0.40(2)  &-2.158(2)  &-1.536( 7)  & 2.70( 3)\\
  0.70 &-0.111(5) &-0.091(1)  & 0.09(2)  &-1.909(2)  &-0.904(11)  & 4.37( 5)\\
  0.60 & 0.006(6) &-0.035(1)  &-0.18(3)  &-1.628(4)  &-0.505( 9)  & 4.88( 5)\\
  0.50 & 0.115(6) &-0.004(1)  &-0.52(3)  &-1.285(4)  &-0.354( 7)  & 4.05( 4)\\
  0.40 & 0.193(5) & 0.023(1)  &-0.74(2)  &-0.891(4)  &-0.223( 4)  & 2.91( 2)\\
\hline
S1 &$\ssfr=1.1$ &r1\\
\hline
  0.40 & 0.192(5) & 0.022(1)  &-0.75(2)  &-0.889(4)  &-0.227( 2)  & 2.93( 2)\\
  0.30 & 0.248(4) & 0.041(1)  &-0.92(2)  &-0.480(5)  &-0.153( 2)  & 1.45( 3)\\
  0.20 & 0.282(5) & 0.053(1)  &-1.01(2)  &-0.128(4)  &-0.105( 1)  & 0.10( 2)\\
\hline
S2 &$\ssfr=1.1$ &r2\\
\hline
  1.10 &-0.214(5) & 0.026(4)  & 1.05(3)  &-3.011(2)  &-3.004( 2)  & 0.03( 1)\\
  1.00 &-0.130(6) & 0.102(4)  & 1.01(3)  &-2.876(2)  &-2.863( 2)  & 0.06( 2)\\
  0.90 &-0.064(3) & 0.039(2)  & 0.45(2)  &-2.693(2)  &-2.329( 5)  & 1.58( 2)\\
  0.80 & 0.003(3) &-0.028(2)  &-0.13(1)  &-2.456(2)  &-1.723( 9)  & 3.18( 4)\\
  0.70 & 0.064(5) &-0.044(1)  &-0.47(2)  &-2.130(4)  &-1.098(13)  & 4.49( 6)\\
  0.60 & 0.127(7) &-0.016(1)  &-0.62(3)  &-1.728(6)  &-0.648(10)  & 4.69( 5)\\
  0.50 & 0.176(5) & 0.013(1)  &-0.71(2)  &-1.255(6)  &-0.420( 9)  & 3.63( 5)\\
  0.40 & 0.230(6) & 0.037(1)  &-0.84(3)  &-0.801(6)  &-0.238( 5)  & 2.45( 4)\\
\hline
S1 &$\ssfr=1.1$ &r2\\
\hline
  0.40 & 0.231(5) & 0.038(1)  &-0.85(2)  &-0.795(5)  &-0.225( 3)  & 2.52( 3)\\
  0.30 & 0.277(5) & 0.055(1)  &-0.98(2)  &-0.407(5)  &-0.153( 2)  & 1.12( 2)\\
  0.20 & 0.294(7) & 0.066(1)  &-1.01(3)  &-0.095(1)  &-0.095( 1)  & 0.00( 1)\\
\hline
S2 &$\ssfr=1.1$ &r3\\
\hline
  1.00 &-0.370(4) &-0.122(3)  & 1.08(2)  &-2.530(2)  &-2.505( 2)  & 0.11( 1)\\
  0.90 &-0.261(5) &-0.039(5)  & 0.96(4)  &-2.360(4)  &-2.346( 4)  & 0.06( 3)\\
  0.80 &-0.161(3) &-0.076(2)  & 0.37(2)  &-2.165(2)  &-1.678( 6)  & 2.11( 3)\\
  0.70 &-0.060(5) &-0.064(1)  &-0.02(2)  &-1.911(4)  &-0.857(10)  & 4.58( 4)\\
  0.60 & 0.037(6) &-0.024(1)  &-0.26(3)  &-1.587(4)  &-0.577(13)  & 4.39( 6)\\
  0.50 & 0.130(5) & 0.014(1)  &-0.50(2)  &-1.239(4)  &-0.370( 8)  & 3.78( 4)\\
  0.40 & 0.218(5) & 0.040(1)  &-0.77(2)  &-0.860(4)  &-0.220( 5)  & 2.78( 3)\\
\hline
S1 &$\ssfr=1.1$ &r3\\
\hline
  0.40 & 0.213(5) & 0.041(1)  &-0.76(2)  &-0.858(4)  &-0.219( 3)  & 2.83( 2)\\
  0.30 & 0.271(4) & 0.056(1)  &-0.95(2)  &-0.484(5)  &-0.144( 2)  & 1.50( 2)\\
  0.20 & 0.301(7) & 0.069(1)  &-1.03(3)  &-0.100(2)  &-0.092( 1)  & 0.04( 1)\\
\hline
\end {longtable}

\end{document}